\documentclass[10pt,letterpaper]{article}
\usepackage{opex3}
\usepackage{amsmath}
\usepackage{cite}

\begin{document}

\title{Route to strong localization of light: the role of disorder}

\author{Diego Molinari$^{1,2}$ and Andrea Fratalocchi$^{1,*}$}
\address{$^1$PRIMALIGHT (www.primalight.org), Faculty of Electrical Engineering; Applied Mathematics and Computational Science, King Abdullah University of Science and Technology (KAUST), Thuwal 23955-6900, Saudi Arabia}
\address{$^2$Dept. of Astronomy, Bologna University, via Ranzani 1, I-40127, Bologna, Italy.}
\email{andrea.fratalocchi@kaust.edu.sa} 


\begin{abstract}
By employing Random Matrix Theory (RMT) and first-principle calculations, we investigated the behavior of Anderson localization in 1D, 2D and 3D systems characterized by a varying disorder. In particular, we considered random binary layer sequences in 1D and structurally disordered photonic crystals in two and three dimensions. We demonstrated the existence of a unique optimal degree of disorder that yields the strongest localization possible. In this regime, localized modes are constituted by defect states, which can show subwavelength confinement properties. These results suggest that disorder offers a new avenue for subwavelength light localization in purely dielectric media.
\end{abstract}




\section{Introduction} 
One of the most fascinating phenomena occurring in disordered systems is  Anderson localization \cite{Anderson57,John87,conti08:_dynam_light_diffus_three_dimen,PhysRevLett.55.2692,ol.34.130,PhysRevLett.55.2696,Wiersma97,chabanov00:_statis_signat_of_photon_local,prb.81.014209,schwartz07:_trans_and_ander_local_in,PhysRevLett.96.063904,PhysRevB.73.245107,scheffold99:_local_or_class_diffus_of_light,PhysRevLett.96.043902,PhysRevLett.53.2169,PhysRevB.47.13120,PhysRevE.51.6301,PhysRevB.60.1555,PhysRevE.73.056616}. This dynamics relies on the restoration of interference effects due to coherent scattering, which alters the eigenmodes of the system from extended to localized and traps the electromagnetic radiation into long-living resonant states. The transition to Anderson localization is normally manifested as the breaking of light diffusion, with the emergence of localized states whose properties are described by a characteristic length ---namely the localization length $l_c$--- that measures the spatial degree of localization achieved through Anderson modes \cite{Sheng}. Diffusion and localization can share an intermediate stage known as weak localization, which acts as a precursor for Anderson localization to settle in \cite{Sheng,PhysRevLett.56.1471}. In the research panorama initiated by Anderson, an open problem is represented by the regime of strong localization of light, where interference effects are expected to be more pronounced \cite{Sheng}. In this situation, a full vectorial solution of Maxwell's equations \cite{conti08:_dynam_light_diffus_three_dimen} can be applied to rigorously study the problem. Among the various open issues, a fundamental question concerns the role of disorder in sustaining strongly localized states, and how to effectively vary the randomness in order to control localized modes and achieve strong localization effects. This problem is essential not only to capture new insights on the dynamics of localization, but also to identify new avenues for the application of disordered systems to the manipulation of light properties.\\
In this Article, we combine Random Matrix Theory (RMT) with massively parallel Finite-Difference Time-Domain (FDTD) simulations, and study the behavior of light localization versus disorder in 1D, 2D and 3D systems. More specifically, we begin with one dimensional media and employ RMT \cite{crisanti93:_produc_of_random_matric} to demonstrate the existence of a unique minimum in $l_c$, as well as the existence of an optimal degree of disorder for localizing light. We then considered 2D and 3D geometries, and evaluated through parallel FDTD simulations the behavior of $l_c$ in the spectral region where the highest density of localized states is expected to be observed. In particular, we considered different realizations of disordered photonic crystals and analyzed the properties of localized states near the band edge of the photonic bandgap, which is expected to support a high density of localized states \cite{John87}. As in the 1D case, we found that the variation of $l_c$ is characterized by a bell-shaped behavior showing a unique minimum associated to the strongest localization possible. For average disorder values different from this optimal condition, strong localization of light is completely lost. This contributes, in part, to explain the elusive character of this phenomenon. Quite remarkably, our analysis reveals that in the condition of maximal localization, electromagnetic modes are constituted by defect states, whose localization length is smaller than one cell of the periodic crystal and has a subwavelength spatial extension, which in both 2D and 3D cases is approximatively $60\%$ smaller than a single internal wavelength in the high index refractive material. This demonstrates that a proper engineering of disorder can be effectively employed for subwavelength control of light within purely dielectrics.\\
\begin{figure}
\centering
\includegraphics[width=8.5cm]{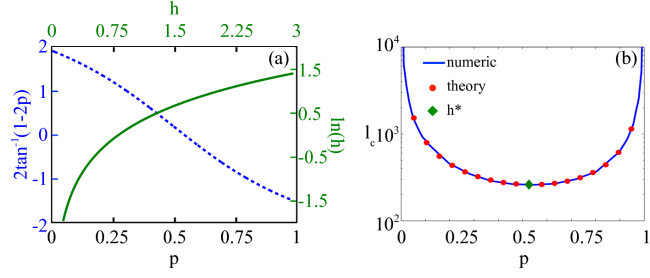}
\caption{(a) Graphical analysis of Eq. (\ref{min}); (b) Localization length $l_c$ vs $p$ for $\gamma_a=\pi/4$, $h_a=h_b=1$ and $\gamma_b=\pi/4+0.2$: numerical evaluation based on the product of $10^6$ matrices (solid line), analytical estimates (circle markers) from Eq. (\ref{sol}) and localization minimum (diamond marker) from Eq. (\ref{min}). 
\label{one}
}
\end{figure}
\section{1D systems: a unique minimum in the localization length}
We begin by considering the $(X,Z)$ propagation of light into a disordered one-dimensional system characterized by $N$ different layers, each possessing a refractive index $n_i$ and thickness $h_i$, which are randomly varying on the $X$ axis. Maxwell's equations are solved by the transfer matrix approach, by looking for solutions of the type $E_y(x,z)=\frac{E_0}{2}\mathcal{U}(x)e^{i\omega t-i\beta z}+\mathrm{c.c.}$ and $H_z(x,z)=\frac{E_0}{2ic\mu_0}\mathcal{V}(x)e^{i\omega t-i\beta z}+\mathrm{c.c.}$,
where $\beta$ is the propagating constant of the mode with frequency $\omega$, $x=X/k_0$, $z=Z/k_0$ dimensionless coordinates ($k_0=2\pi/\lambda$) and $\mathcal{U}$, $\mathcal{V}$ dimensionless fields with $E_0$ is an arbitrary electric field constant. Input $\mathcal{U}_0=\mathcal{U}(0), \mathcal{V}_0=\mathcal{V}(0)$ and output $\mathcal{U}_l=\mathcal{U}(L), \mathcal{V}_l=\mathcal{V}(L)$ ($L$ is the sample length along $x$) are related by the transfer matrix \cite{souk} $\mathbf{M}=\prod_{j=1}^N\mathbf{M}_j$,
with:
\begin{eqnarray}
\label{mj}
\mathbf{M}_j=\begin{bmatrix}\cos{\gamma_jh_j} &-\frac{\sin{\gamma_j h_j}}{\gamma_j}\\\gamma_j\sin{\gamma_j h_j} &\cos{\gamma_jh_j}\end{bmatrix},
\end{eqnarray}
with $\gamma_j=\sqrt{n_j^2-n_\beta^2}$ and $n_\beta=\beta/k_0$. The matrix $\mathbf{M}_j$ belongs to the Lie group \emph{SL(2,R)} and provides a non orthogonal rotation to the input fields $\mathcal{U}_0$ and $\mathcal{V}_0$. According to Furstenberg theorem \cite{crisanti93:_produc_of_random_matric}, when the elements of $\mathbf{M}_j$ are random variables, the following limit holds: $\lim_{N\rightarrow\infty}\langle \log Tr(\mathbf{M})\rangle=\delta>0$ [$\langle...\rangle$ denotes an ensemble average]
and all modes get exponentially localized as $\mathcal{U},\mathcal{V}\propto e^{-\delta |x|}$, where $\delta$ is the positive Lyapunov exponent of the random matrix $\mathbf{M}$ \cite{crisanti93:_produc_of_random_matric}.
The localization length $l_c$ can be directly computed from the Lyapunov exponent and reads $l_c=\delta^{-1}$. We focused our analysis in the important physical situation where the disordered material is composed by a random binary sequence of two different media, each characterized by a diverse index $n_i$ and/or thickness $h_i$, with $i\in[1,2]$. These define two different transfer matrices $\mathbf{M}_{1,2}$, which we labeled $\mathbf{A}$ and $\mathbf{B}$ for notation simplicity. Random deviates are chosen from a binary probability density characterized by two discrete values: $p$ and $q=1-p$, which represent the probability to extract $n_1$ ($h_1$) or $n_2$ ($h_2$) values, respectively. A binary superlattice is not only physically interesting \emph{per se} \cite{PhysRevLett.101.183901,PhysRevLett.99.063905}, but it is also fundamental to understand the behavior of the localization length $l_c$ in higher dimensions. In 2D and 3D systems, in fact, the majority of the experiments is conducted in samples showing the random aggregation of two different materials such as, e.g., scatterers hosted in a liquid matrix (a photonic colloid) or disordered solids where the second medium is composed by air holes \cite{PhysRevLett.55.2692,PhysRevLett.55.2696,Wiersma97,schwartz07:_trans_and_ander_local_in,apl}. In all these cases, a 1D representation of the dynamics along a specific direction of propagation, or parallel to a light ray that is randomly scattering inside the sample, would be equivalent to that of a random sequence of two different layers.
We analyzed the dynamics of $l_c$ versus disorder by applying a fully analytic theory based on the Microcanonical method. The method is well documented in \cite{PhysRevLett.62.695,crisanti93:_produc_of_random_matric} and in this paper we show only the main results achieved with this approach. We focused on the following general case:
\begin{equation}
\mathbf{M}=\mathbf{M}_0+\Delta\mathbf{M},
\end{equation}
with $\mathbf{M}_0$ a constant matrix and $\Delta\mathbf{M}$ a random term. In order to evaluate the behavior of $l_c$, we calculated the generalized Lyapunov exponent $\mathcal{L}(1)=\lim_{N\rightarrow\infty}\log\langle Tr(\mathbf{M})\rangle=\tilde{\delta}$ that, apart from negligible fluctuations, coincides with $\delta$ \cite{PhysRevLett.62.695,crisanti93:_produc_of_random_matric}. The ensemble average can be expressed as follows:
\begin{equation}
\langle Tr(\mathbf{M})\rangle=Tr(\mathbf{M}_0)+\langle Tr(\Delta\mathbf{M})\rangle,
\end{equation}
with $Tr(\mathbf{M}_0)$ a constant value and $\langle Tr(\Delta\mathbf{M})\rangle$ expressed by the following Cauchy integral \cite{crisanti93:_produc_of_random_matric}:
\begin{eqnarray}
\label{tr}
\langle Tr(\Delta\mathbf{M})\rangle=\begin{pmatrix}N\\N'\end{pmatrix}^{-1}\frac{1}{2\pi i}\int_\mathcal{C}\mathbf{Q}(z)e^{N\ln\frac{\gamma}{z^(1-p)}}dz
\end{eqnarray}
where $\mathbf{Q}(z)$ is an inessential nonextensive matrix of order $O(1)$, $\mathcal{C}$ an arbitrary contour enclosing $z=0$ in the complex plane $z$, $N'=(1-p)N$, $\gamma$ the largest eigenvalue of the matrix $\mathbf{A}+z\mathbf{B}$ and $p$ the probability of extracting the matrix $\Delta\mathbf{M}=\mathbf{A}$ from the binary ensemble $\Delta\mathbf{M}\in[\mathbf{A},\mathbf{B}]$. Equation (\ref{tr}) can be evaluated with the saddle point method and, after straightforward algebra, we obtain:
\begin{eqnarray}
\label{sol}
&\delta(p)=\ln\bigg[ \bigg(\frac{1-p}{h}\bigg)^{1-p}p^p\gamma\bigg], &\frac{\partial\ln\gamma}{\partial\ln z}\bigg|_{z=h}=1-p,
\end{eqnarray}
with $h$ a complex value that satisfies the saddle point equation (\ref{sol}), which can be employed to study the behavior of the localization length versus disorder. The latter can be measured by the probability $p$ of extracting the matrix $\mathbf{A}$ from the binary ensemble. Values of $p=0$ or $p=1$, in particular, describe an ordered system composed by a constant configuration of $\mathbf{A}$ or $\mathbf{B}$ matrices, while $p\in(0,1)$ yields different disordered sequences of $\mathbf{A}$ and $\mathbf{B}$. By applying a derivative on $p$, we get $\partial\delta/\partial p=2\tan^{-1}(1-2p)-\ln h$, which shows a single zero at $p^*$:
\begin{equation}
\label{min}
2\tan^{-1}(1-2p^*)=\ln h(p^*),
\end{equation}
and supports a unique extremum in the localization length $l_c$, as shown graphically in Fig. \ref{one}(a). As a result, there exists a single optimal degree of disorder that yields the strongest localization in the material. We verified this theoretical prediction by performing a series of numerical simulations. Without any loss of generality, we set the arbitrary term $\mathbf{M}_0$ to $\mathbf{M}_0=0$. We then considered two matrices $\mathbf{A}$ and $\mathbf{B}$ with $\gamma_a=\pi/4$, $\gamma_b=\gamma_a+0.2$ and $h_a=h_b=1$ (since $h_j$ appears always in the factor $\gamma_j h_j$, a random variation of $\gamma_j$ is enough to have a binary random sequence) and compared the localization length $l_c$ with Eqs. (\ref{sol})-(\ref{min}). An excellent level of agreement is found between theory and numerical results [Fig. \ref{one}(b)], with the localization length showing a single minimum as predicted by Eq. (\ref{min}).\\
\begin{figure}
\centering
\includegraphics[width=12cm]{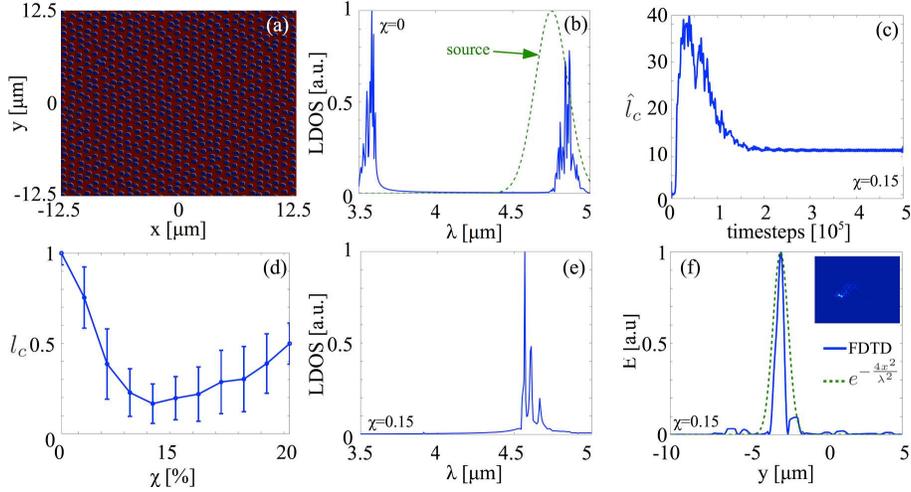}
\caption{Summary of two dimensional FDTD simulations. (a) Refractive index distribution for $\chi=0.15$; (b) local density of states LDOS (solid line) for the ordered crystal at $\chi=0$ and spectrum of the input source used in FDTD simulations (dashed line); (c)
single localization length $\hat{l}_c$ for a sample with $\chi=0.15$; (d) normalized averaged localization length $l_c$ versus disorder strength $\chi$: (e)-(f) strongest localization regime for $\chi=0.15$: (e) LDOS and (f) energy section along $y$ (solid line) compared to a gaussian spot (dashed line) whose size $s=2\omega_y$ (being $\omega_y$ the waist) corresponds to one internal wavelength $\lambda$ in Silicon. The inset of (f) reports the spatial energy distribution $\mathcal{E}(x,y)$ in the strongest localized case. 
\label{two}
}
\end{figure}
\section{2D and 3D media: defects states and subwavelength effects}
\begin{figure}
\centering
\includegraphics[width=10cm]{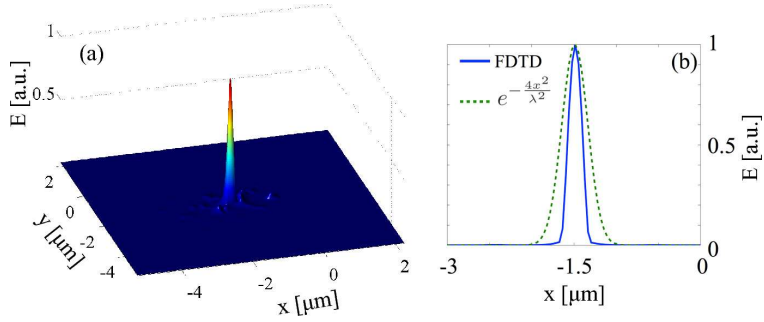}
\caption{3D subwavelength localization for $\chi=0.05$: (a) spatial $(x,y)$ distribution of the electromagnetic energy $\mathcal{E}(x,y,z)$ calculated at $z=-2\mu$m; (b) Energy $\mathcal{E}$ profile along $x$ for $z=-2\mu m, y=-1.5\mu m$ (solid line) compared to an equivalent gaussian spot of size equal to one internal wavelength $\lambda=\lambda_0/n$ (dashed line). 
\label{four}
}
\end{figure}

We studied the behavior of higher dimensional systems by resorting to FDTD simulations, and performed a massively parallel simulation campaign (about 10M of single-cpu hours) on a variety of experimentally feasible 2D and 3D structures. As a measure of $l_c$, we employed the Inverse Participation Ratio (IPR):
\begin{equation}
\label{ipr}
l_c=\langle\hat{l}_c\rangle=\bigg\langle\frac{\big(\int_V \mathrm{d\mathbf{x}}\mathcal{E}\big)^2}{V\int_V\mathrm{d\mathbf{x}}\mathcal{E}^2}\bigg\rangle,
\end{equation}
with $\hat{l}_c$ the IPR of a single realization of disorder, $V$ the system volume and $\mathcal{E}=\frac{1}{2}(\mathbf{E}\cdot\mathbf{D}+\mathbf{H}\cdot\mathbf{B})$ the electromagnetic energy density. The operator $\langle ... \rangle$ represents an average over realizations.
Random samples are constructed by disordering a Photonic Crystal (PhC) of air holes on a Silicon (Si) substrate. In particular, we considered positional disorder of the type $\mathbf{x}=\mathbf{x}_0+\Delta$, with $\Delta$ being a two/three element random vector whose values $\Delta_i$ are given by a normal distribution with $\langle \Delta_i\rangle=0$ and standard deviation $\sqrt{\langle\Delta_i^2 \rangle}=\chi$. In order to prevent folding effects, we constrained the global shift of each hole to be lower than the lattice constant of the PhC, and chose the standard deviation $\chi$ of disorder accordingly. In our FDTD simulations, in order to get accurate results, we adopted a spatial discretization with at least $20$ internal points per wavelengths in the high-index (Si) material.\\
Two dimensional samples are constructed from a triangular lattice, $66 \mu$m $\times 66\mu$m wide along $x$ and $y$, characterized by air holes of radius $r=0.3 \mu$m. Figure \ref{two}(a) shows the refractive index distribution of a single realization for $\chi=0.15$. In our simulation campaign, we considered on average 20 disorder realizations for each value of $\chi$, which we have verified is sufficiently high to yield a good convergence of the averaged observables. In order to maximize the probability of observing localization effects, we employed a wideband gaussian source with FWHM=$250$nm centered at $\lambda_0=4.8\mu m$ [Fig. \ref{two}(b) dashed line], whose wavelength is close to one of the gap edges of the  photonic crystal, as shown by the Local Density Of States (LDOS) of the system for $\chi=0$ [Fig. \ref{two}(b) solid line]. In each simulation, we launched the gaussian source and then monitored the evolution of $\hat{l}_c(t)$ over time $t$ [Fig. \ref{two}(c)]. After an initial transient of $\approx 4\cdot 10^5$ timesteps, the source propagated over the system and the localization length $\hat{l}_c$ reached a steady state, characterized by the electromagnetic modes excited in the system after the interaction with the input wave [Fig. \ref{two}(c)]. The behavior of $l_c$, after averaging, is qualitatively identical to the one dimensional case, with a single minimum attained in the dynamics [Fig. \ref{two}(d)]. 
\begin{figure}
\centering
\includegraphics[width=8.5cm]{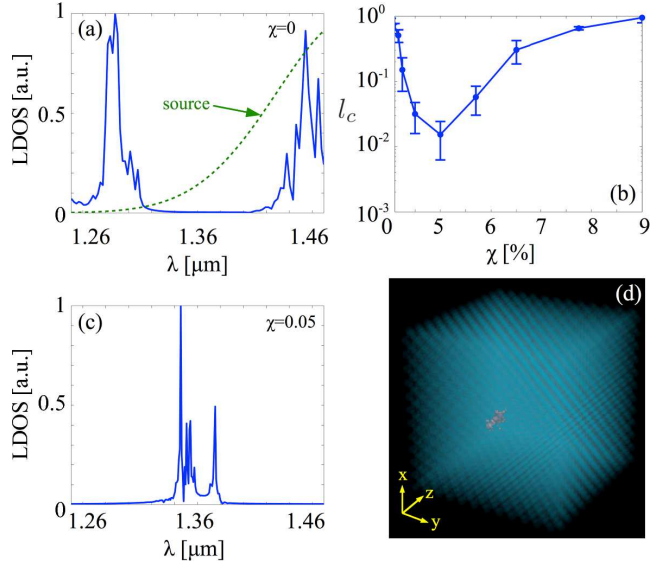}
\caption{(a) LDOS for the ordered crystal (solid line) and input source spectrum (dashed line) used in 3D FDTD experiments; (b) normalized localization length $l_c$ versus $\chi$; (c)-(d) strongest localization regime for $\chi=0.05$: (c) LDOS and (d) refractive index (semitransparent volume density plot) with electromagnetic energy (isosurface plot) distribution.
\label{three}
}
\end{figure}
In order to investigate the properties of light in the regime of strong localization, we begin by calculating the LDOS for the strongest localization, which is attained for $\chi=0.15$ [Fig. \ref{two}(e)]. As seen in Fig. \ref{two}, electromagnetic states are composed by a few eigenfrequencies located inside the badgap of the photonic crystal, and they effectively behave as single-defect states. As a result, their spatial localization is nor originated by Anderson interference effects, neither by multiple impurity scattering, but conversely by the localization properties of the electromagnetic bandgap. The consequences of this dynamics are quite interesting, as shown in Fig. \ref{two}(f), which displays the spatial distribution of the electromagnetic energy $\mathcal{E}$ for one realization at $\chi=0.15$. As seen, the electromagnetic radiation is able to reach a subwavelength confinement. The FWHM of the energy spot, in fact, is calculated to be $FWHM=0.58\mu$m, which is $60\%$ smaller than one internal wavelength $\lambda=\lambda_0/n=1.38\mu$m, where $n$ is the refractive index of Si. It is worthwhile remarking that such a subwavelength confinement can be only observed in the narrow range of $\chi$ that yields a strong localization regime, i.e., for $\chi\in[0.13,0.18]$. As seen in Fig. \ref{two}(d), in fact, a variation of a few percent of average disorder from its optimal condition yields an increase of the localization length, with the vanishing of subwavelength confinement.\\
We completed our theoretical analysis by the study of 3D systems. FDTD simulations have been carried out by disordering a $18 \mu m\times 18\mu m\times 18\mu m$ fcc lattice of air holes (with radius $r=0.35 \mu$m) in silicon, and by employing a Gaussian input source with $\lambda_0=1.5 \mu$m and $\mathrm{FWHM}=100$nm. As in the 2D case, we have centered the source at one edge of the photonic crystal gap in order to maximize the probability of observing localized states [Fig. \ref{three}(a) dashed line]. Figure \ref{three}(b)-(d) summarizes the results arising from numerical simulations. The behavior of the localization length $l_c$ is qualitatively identical to the 2D case [Fig. \ref{two}(d)], with a unique minimum observed as $\chi$ is varied. Quantitatively, the strongest localization is attained at $\chi=0.05$ [Fig. \ref{three}(b)], which corresponds to a small variation of disorder of only $5\%$. In this regime, as for the two-dimensional case, strongly-localized states are characterized by defect-states originated in the electromagnetic bandgap, as observed by the LDOS reported in Fig. \ref{three}(c). In order to further investigate the degree of spatial confinement achieved in the strong localization regime, we considered a single realization for $\chi=0.05$ and evaluated the FWHM of the hot energy spot [Fig. \ref{four}(a)-(b)], obtaining a value $FWHM=0.19\mu$m, which is $\approx 56\%$ smaller than one internal wavelength $\lambda_0/n=0.43\mu$m in silicon. The three dimensional system therefore shows the same qualitatively scenario of the two dimensional case, with comparable subwavelength confinement properties ---i.e., $60\%$ in 2D vs $56\%$ in 3D--- although it offers a much narrower disorder window for achieving strong localization effects, as observed by comparing Fig. \ref{three}(b) with Fig. \ref{two}(d).
\section{Discussion and conclusions} 
We investigated the role of disorder in sustaining strong localization of light in 1D, 2D and three-dimensional systems. In one-dimensional media, we applied a Random matrix analysis and demonstrated the existence of a unique minimum that yields the strongest localization possible. A set of numerical simulation has then been employed to verify this result. In 2D and 3D media, conversely, we tackled the problem by an extensive campaign of FDTD simulations, which kept into account the vectorial nature of Maxwell's equations. More specifically, we considered a series of samples characterized by structurally disordered photonic crystals. The behavior of the localization length $l_c$ in these systems, as obtained by FDTD simulations, appears qualitatively identical to the one dimensional case, with the manifestation of a single minimum that sustains the strongest localization of light possible. From a more quantitatively perspective, in both 2D and 3D samples, FDTD simulations predicts the possibility of observing strong localization effects for a relatively small average disorder : $\chi=15\%$ in two dimensions and $\chi=5\%$ in the 3D case. In addition, the numerical analysis shows that a small variation of disorder from the optimal condition leads to a complete vanishing of the strong character of localized modes, with the localization length sharply increasing to the initial value pertaining to the unperturbed crystal. The average disorder range for the observation of strong localization effects is quite narrow: approximatively $5\%$ in 2D and only $1\%$ in three dimensions. These results contribute, to a certain extent, to also explain the elusive nature of this phenomenon.\\An interpretation for the qualitatively similar behavior of $l_c$ versus disorder in 1D, 2D and 3D systems can be drawn by generalizing the one dimensional RMT analysis to higher manifolds. In 2D and 3D cases, in fact, we can fold the photons dynamics over the scattering trajectory of light rays. A single ray, in the most common configuration where only two materials are employed (i.e., a suspension of particles in a host matrix or air holes on a disordered substrate \cite{apl}), undergoes a chaotic motion inside the sample and practically experiences a random succession of two layers. The presence of a higher dimensional manifold allows a richer dynamics of the evolution of light paths if compared to the one dimensional case, which might lead to a different scenario for light localization (e.g., mobility edge, marginal localization, coexistence of localized and extended states, ...). However, among all the various paths available, we can intuitively expect that the variation of disorder can lead to the formation of a minimum in $l_c$ when, on average, light rays scatter along trajectories whose disorder is optimal for their one dimensional evolution. Regardless of the presence of a higher manifold for the dynamics of light, we can therefore qualitatively expect the appearance of a single minimum, if the average disorder is able to optimally localize the dynamics of light rays.\\As a final step of our theoretical analysis, we investigated the nature of localized modes in the strongest localization regime. Calculation of the local density of states revealed that in this condition light localization is sustained by defect states originated by a few eigenfrequencies inside the electromagnetic band gap. This results in a very strong spatial confinement for light, due to the excitation of such disordered-induced impurity modes. From a physical perspective, such a strong confinement can be explained by considering that a defect mode results from the interference of a large number of plane waves that, being localized in the spectral bandgap, cannot couple to any extended state of the structure and are forced to interfere at every wavevector $\mathbf{k}$. The ideal case of an infinite sum of plane waves, which coherently interfere at every angle, yields the smallest spot achievable, i.e.,  a dirac-delta $\delta$ function. Conversely, the opposite case of an incoherent summation of plane waves originates the well-known Rayleigh distribution (see, e.g., \cite{haake01:_quant_signat_chaos}) that does not show any hot-spot with subwavelength confinement. Our analysis shows that suitable values the disorder support a coherent combination of plane waves that lives in between these two extrema ---i.e., a full coherent combination and a totally incoherent interference---, which results into a subwavelength spatial confinement for light. A proper engineering of disorder can therefore be employed as an avenue for subwavelength light localization, where extremely narrow hot energy spot can be generated to highly enhance absorption and/or emission properties of purely dielectric materials. This is expected to open new applications in the field of high power lasers, sensing and energy harvesting devices. 

\section*{Acknowledgments}
For computer time, this research used the resources of the 
Supercomputing Laboratory at King Abdullah University of Science \& Technology (KAUST) in 
Thuwal, Saudi Arabia. Numerical simulations have been performed with our NANOCPP code, which is an homemade, highly scalable 2D/3D FDTD code expressively developed for large scale parallel simulations of disordered materials. D. Molinari acknowledges partial support from PRIN MIUR 2009. The authors thank G. Ruocco and P. de Bernardis for fruitful discussions.

\end{document}